# An explanation for the gap in the Gaia HRD for M-dwarfs

James MacDonald & John Gizis
Department of Physics and Astronomy, University of Delaware, Newark, DE 19716, USA

**Abstract**

We show that the recently discovered narrow gap in the Gaia Hertzsprung - Russell Diagram near $M_G =10$ can be explained by standard stellar evolution models and results from a dip in the luminosity function associated with mixing of $^3$He during merger of envelope and core convection zones that occurs for a narrow range of masses.

**Introduction**

Based on measurements presented in Gaia Data Release 2 (Gaia Collaboration et al. 2016, 2018), Jao et al. (2018) have reported their discovery of a narrow gap in the main sequence on the Hertzsprung - Russell Diagram (HRD) near $M_G =10$. The gap is near spectral type M3V where M dwarf stars are thought to transition from partially to fully convective, and Jao et al. propose that the gap is linked to the onset of full convection in M dwarfs.

Here we show that the gap is a feature of existing models and is due to a dip in the luminosity function associated with mixing of $^3$He during merger of envelope and core convection zones. The narrowness of the gap is a result of the small mass range over which this merger can occur.

**Modeling description**

We have used our fully – implicit stellar evolution code to follow models of masses 0.300 to 0.380 $M_\odot$ in increments of 0.005 $M_\odot$ from the pre-main sequence to age equal to that of the Universe. Our models use outer boundary conditions based on the BT-Settl atmosphere models (Allard, Homeier & Freytag 2012, Allard et al. 2012, Rajpurohit et al. 2013). Specifically, we use the temperature and pressure from the atmosphere models at optical depth $10^3$ to set the outer boundary condition. For the interior, we use the OPAL radiative opacities (Iglesias & Rogers 1996) for temperatures greater than 20000 K, and the Ferguson et al. (2005) and Freedman et al. (2008, 2014) radiative opacities at lower temperatures. The Potekhin conductive opacities are used (Potekhin, et al. 1999; Gnedin, Yakovlev & Potekhin 2001; Shternin & Yakovlev 2006; Cassisi et al. 2007). The convective flux is calculated using Böhm-Vitense (1958) mixing length theory. The mixing length ratio, $\alpha = 1$. Convective mixing is treated as a diffusion process with the diffusion coefficient determined from mixing length theory. This approach avoids the problem with instantaneous mixing discussed by Chabrier & Baraffe (1997) that occurs because the deuterium burning time scale can be shorter than the convective turnover time scale. The fully – implicit nature of our code also prevents the convective kissing instability discovered and described by van Saders & Pinsonneault (2012).

All the nuclear reaction rates relevant to low mass stars are from Angulo et al. (1999). The treatment of electron screening is based on Potekhin & Chabrier (2012). In a strongly coupled plasma mixture, the free energy is accurately determined by the linear mixing rule. The enhancement factor for a nuclear reaction is then easily determined from the difference in free energy before and after the reaction

(DeWitt, Graboske & Cooper 1973; Saumon et al 1996). For this regime, we determine the free energy of the one component plasma with rigid electron background and the electron polarization contribution to the free energy using results in Potekhin & Chabrier (2000). In the Debye-Hückel limit of weak coupling, where the linear mixing rule is not valid, the appropriate screening factor is that determined by Salpeter (1954). To bridge the gap between the weak and strong screening limits, we use the interpolation scheme of Salpeter & Van Horn (1969). The equation of state (EOS) is that of Saumon, Chabrier & van Horn (1995). Since this EOS is for H – He mixtures only, we use the additive volume rule (Fontaine, Graboske & van Horn 1977) to add the contribution from heavy elements. The EOS for the heavy elements is based on free energy minimization and has its origins in the Eggleton, Flannery & Faulkner (1973) EOS. However, we include many additional physical processes. We include all ionization stages of C, N, O, Ne, Mg, Si, Fe with each ionization stage assumed to be its ground state. Coulomb and quantum corrections are included as described in Iben, Fujimoto & MacDonald (1992). A hard sphere free energy term is used to model pressure ionization. Electric micro-field (Stark) contributions are treated as in equation 33 of Eggleton et al. (1973) but with the right hand side multiplied by 0.6.

For the composition of the interior, we use the same heavy element abundance, $Z$, as used for the BT-Settl atmosphere calculations, $Z = 0.0153$ (Caffau et al. 2011).

**The luminosity function**

We have created a luminosity function, shown in figure 1, by randomly sampling mass between 0.30 and 0.38 $M_\odot$ and randomly sampling age between 0 and 8.8 Gyr. Bolometric corrections are obtained from the online BT-Settl atmosphere tables[1]. For the initial mass function (IMF), we use a power law fit to the Parravano, McKee & Hollenbach (2011) IMF over the relevant mass interval. Since the M dwarfs observed by Gaia have distances less than the scale height of the Galactic thin disk, the upper age limit is taken to be the age of the thin disk (del Peloso et al. 2005). A dip in the luminosity function is clearly seen near log $L/L_\odot$ = -1.84, which corresponds to $M_K$ = 6.71 in good agreement with the location of the gap shown by Jao et al. (2018) in their figure 6.

The origin of the dip can be discerned from figure 2, in which we plot isochrones in the luminosity versus mass diagram. The isochrone 'pinch' extending from near 0.33 $M_\odot$ to near 0.35 $M_\odot$ is a result of a faster change in luminosity than at lower and higher masses. This leads to a lower probability of finding a star at the pinch luminosities, resulting in a dip in the luminosity function over the interval log $L/L_\odot$ ~ -1.85 to log $L/L_\odot$ ~ -1.82 ($M_K$ ~6.74 to ~6.66).

---

[1] https://phoenix.ens-lyon.fr/Grids/BT-Settl/CIFIST2011/

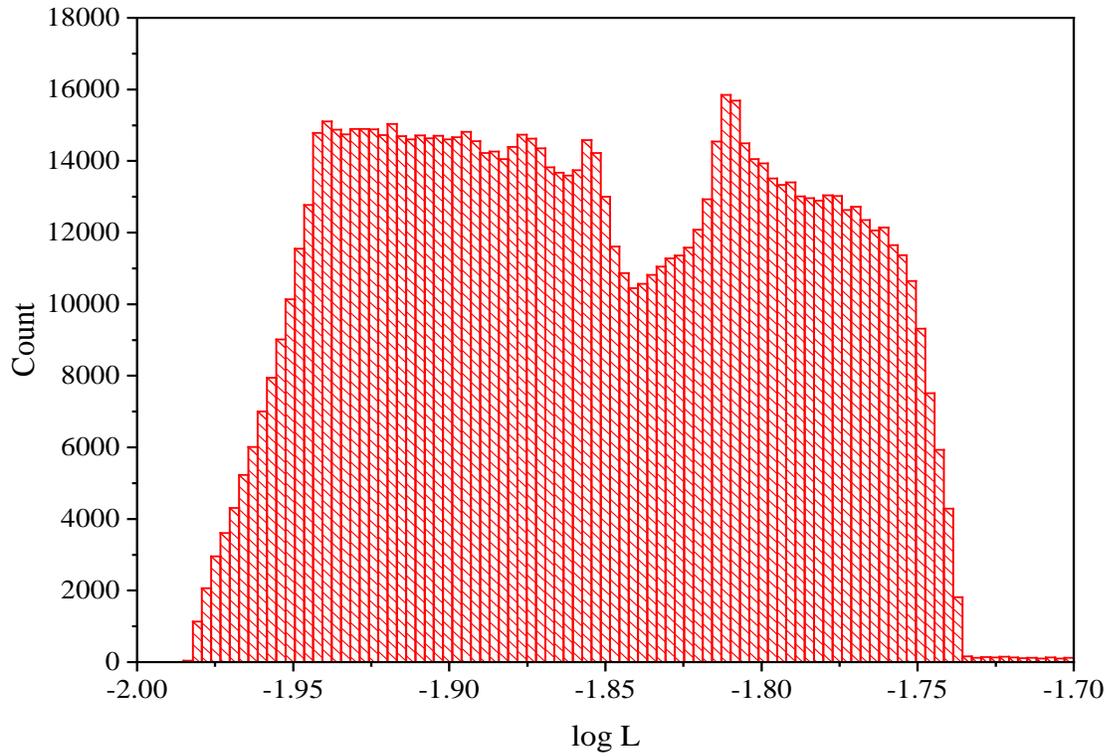

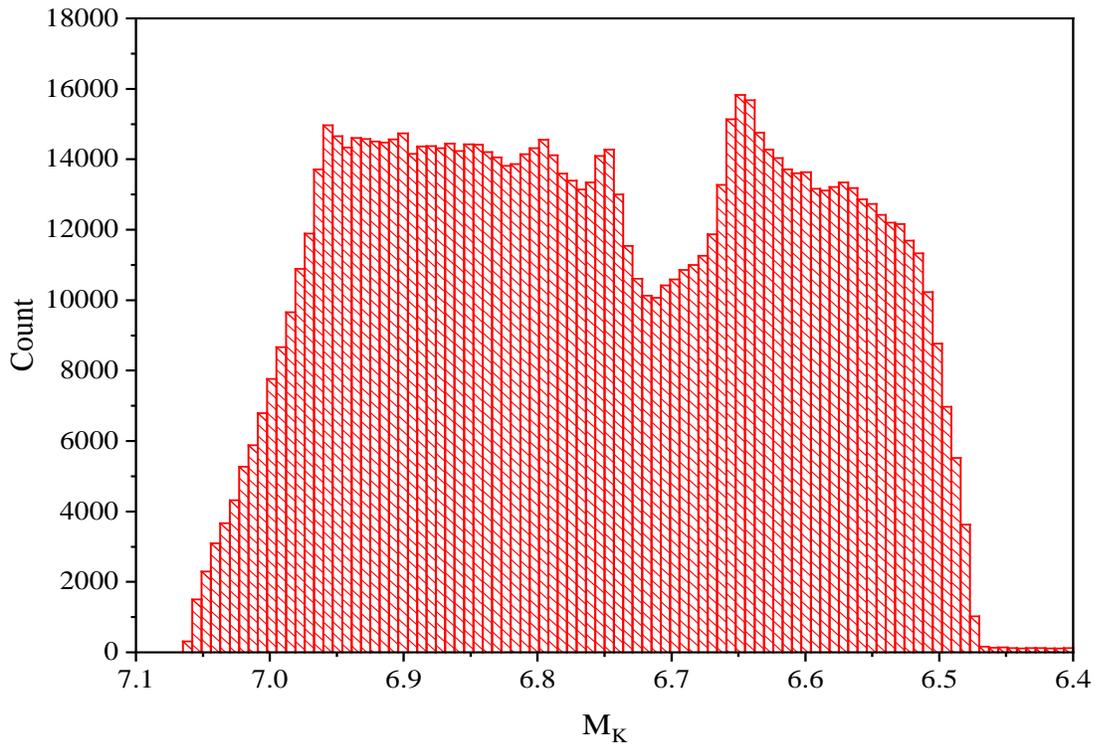

Figure 1. Synthetic luminosity function. The top panel shows a histogram of the luminosity values found by randomly sampling stellar mass between 0.30 and 0.38 $M_\odot$ and stellar age between 0 and 8.8 Gyr. The bottom panel is the same except for the absolute 2MASS $K_s$-band magnitude.

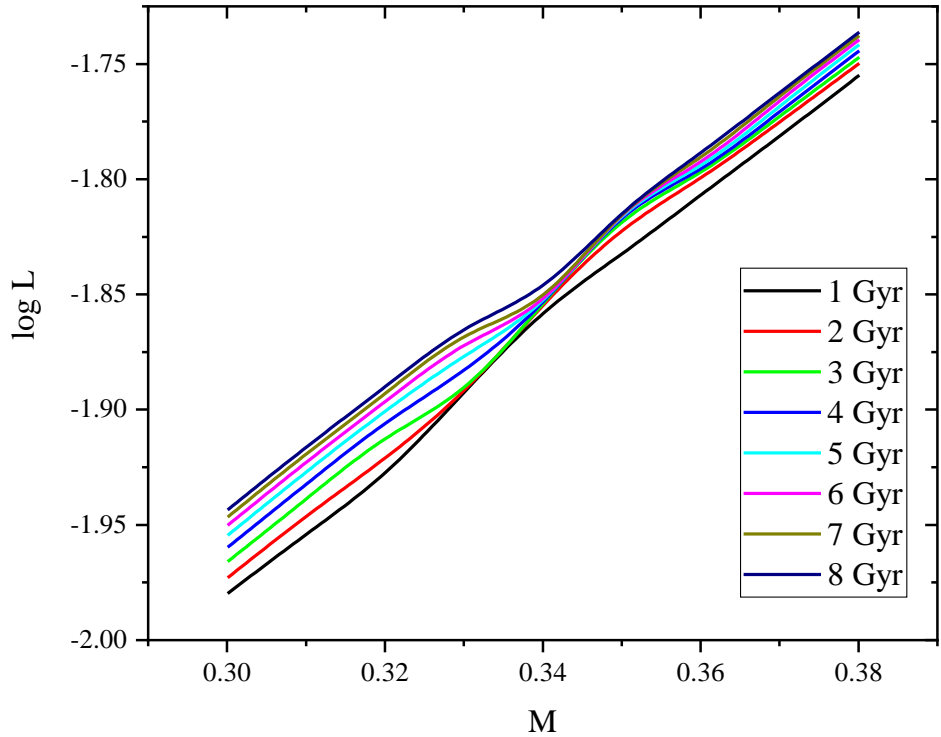

Figure 2. Isochrones in the mass – log luminosity diagram. Mass and luminosity are in solar units.

The physical process responsible for the isochrone pinch involves the complex interplay production of $^3$He and its transport by convective motions, as well as the differences in the details of the evolution of fully convective stars and those that have a radiative zone separating a convective envelope from a convective core. The transition between these behaviors for main sequence stars occurs over the narrow mass range from 0.315 to 0.345 M$_\odot$. In this mass range, before the convective core and envelope merger, the convective core contains a smaller amount of mass than the convective envelope. In the initial main sequence evolution of low mass stars that have a radiative layer between core and envelope convection zone, the core $^3$He abundance increases until it reaches a quasi-equilibrium value. The envelope $^3$He abundance is initially less than the core equilibrium value but also increases with time and, depending on the mass, can exceed the core equilibrium value. If the core and envelope convection zones then merge, the central $^3$He abundance increases causing an increase in luminosity. Figure 3 shows how the luminosity and $^3$He mass fraction at the stellar centre evolve with time for models of masses that span the range in which core – envelope merger occurs.

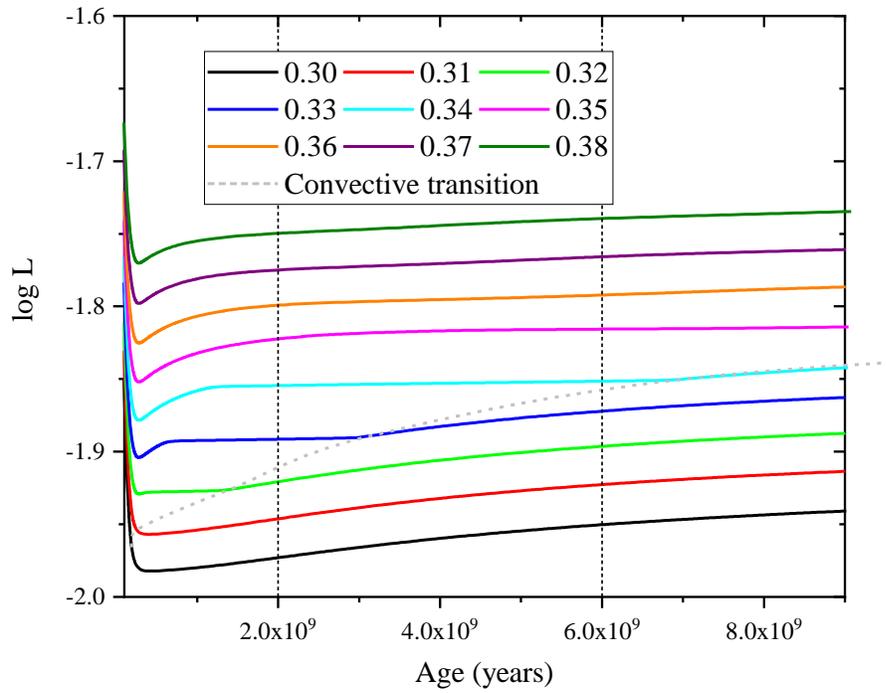

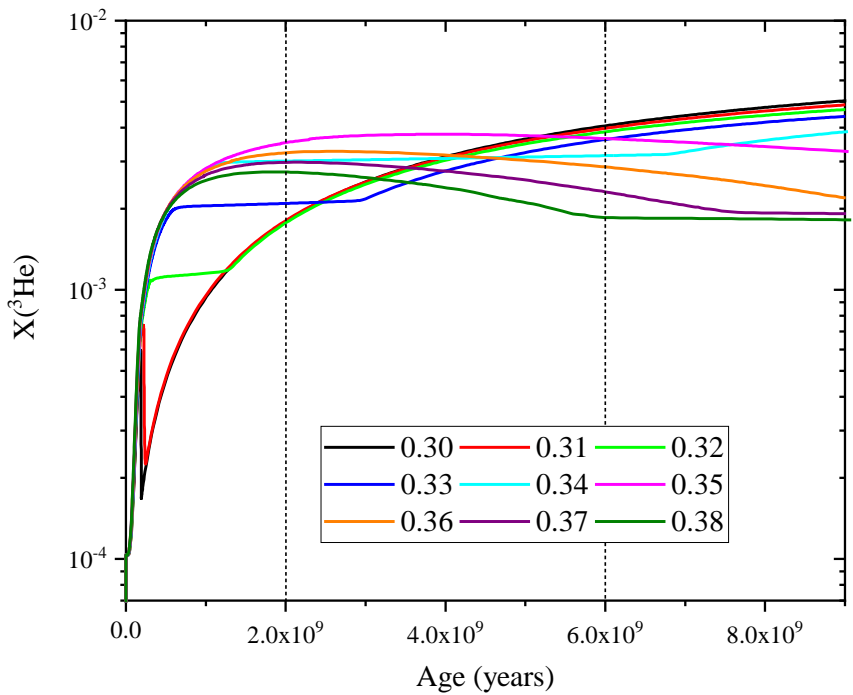

Figure 3. The upper panel shows the evolution of stellar luminosity for models of mass in solar units indicated in the legend. The broken grey line shows where merger of core and envelope convection zones occur. The lower panel shows the evolution of the central value of the $^3$He mass fraction. The vertical dashed lines are guides to facilitate comparison with figure 2.

For the 0.30 and 0.31$M_\odot$ models, the star becomes fully convective before the main sequence phase. The luminosity increases steadily and smoothly because $^3$He is produced in the outer parts of the nuclear energy generation region and is convectively mixed to the centre. The models more massive than 0.35 $M_\odot$ do not experience core – envelope merger and the luminosity remains fairly constant or increases slowly during the main sequence phase. In the intermediate mass range, the luminosity remains fairly constant until merger occurs, after which the luminosity steadily increases for the same reasons as for the lower mass models.

**Discussion**

We have presented a relatively simple explanation, based on current stellar evolutionary models, for the presence of a gap in Gaia HRD for M-dwarfs discovered by Jao et al. (2018). Our calculations are only for solar composition models and it remains to be seen whether adding models of other compositions can diminish the luminosity function dip.

We have not considered the effects of magnetic fields. It has been established that magnetic inhibition of convection (Mullan & MacDonald 2001; Feiden & Chaboyer 2012) and/or the presence of dark stellar spots (Spruit & Weiss 1986) results in main sequence models having larger radii and lower luminosity than non-magnetic models of the same mass. In general, the strongest magnetic fields in M-dwarfs are found in those that are rapidly rotating and fully convective (Shulyak et al. 2017). Thus if a large enough fraction of fully convective M-dwarfs are rapid rotators (rotation periods less than a few days according to Shulyak et al.), this could also give or enhance a dip in the luminosity function near the transition to fully-convective.

**Acknowledgements**

We thank the referee, Leo Girardi, for suggestions for improvements to the manuscript and also for a very quick report, and Wei-Chun Jao for helpful suggestions.